\documentclass[aps,prl,preprint,superscriptaddress,amsmath]{revtex4}

\renewcommand\vec[1]{{\bf #1}}

\begin{document}

\title{{\it Ab initio} calculation of the binding energy of impurities in semiconductors: Application to Si nanowires}

\author{Y.~M. Niquet}
\email{yniquet@cea.fr}
\affiliation{CEA-UJF, INAC, SP2M/L\_Sim, 38054 Grenoble Cedex 9, France}
\author{L. Genovese}
\affiliation{European Synchrotron Radiation Facility, 6 rue Horowitz, BP 220, 38043 Grenoble, France}
\author{C. Delerue}
\affiliation{Institut d'\'Electronique, de Micro-\'electronique et de Nanotechnologie (UMR CNRS 8520), D\'epartement ISEN, 41 boulevard Vauban, F-59046 Lille Cedex, France}
\author{T. Deutsch}
\affiliation{CEA-UJF, INAC, SP2M/L\_Sim, 38054 Grenoble Cedex 9, France}

\date{\today}

\begin{abstract}
We discuss the binding energy $E_b$ of impurities in semiconductors within density functional theory (DFT) and the $GW$ approximation, focusing on donors in nanowires as an example. We show that DFT succeeds in the calculation of $E_b$ from the Kohn-Sham (KS) hamiltonian of the ionized impurity, but fails in the calculation of $E_b$ from the KS hamiltonian of the neutral impurity, as it misses most of the interaction of the bound electron with the surface polarization charges of the donor. We trace this deficiency back to the lack of {\it screened} exchange in the present functionals.
\end{abstract}

\maketitle

The binding energy $E_b$ of donors and acceptors is a key quantity in semiconductor physics because it determines the doping efficiency. In semiconductor nanostructures for example, confinement and electrostatics tend to shift the impurity levels deeper in the gap, which decreases dopant activity \cite{Bjork09,Yoon09}. Therefore, {\it ab initio} calculations of impurity binding energies are highly desirable to assess the performances of ultimate nanodevices. Besides, donors and acceptors are the prototypes of charged defects in semiconductors, and a fundamental understanding of the strengths and weaknesses of present {\it ab initio} approaches such as density functional theory (DFT) and the $GW$ approximation \cite{Hedin69,Onida02} would open the way to a more accurate modeling of complex defects.

So far, the calculation of $E_b$ in bulk semiconductors has been possible only with semi-empirical methods \cite{Kohn55,Martins02}. However, calculations based on DFT have become practicable in ultimate nanostructures with a smaller number of atoms. Recently, the case of donors in Si nanowires (Si NWs) has been adressed with both semi-empirical methods and DFT, with contradictory results. Tight-binding \cite{Diarra07,Diarra08} and effective mass calculations \cite{Li08}, supported by experiments \cite{Bjork09,Yoon09}, indeed suggest that $E_b$ increases as $1/R$ with decreasing wire radius $R$, due to the interaction of the bound electron with the surface polarization (or ``image'') charges of the impurity, resulting in a significant decrease of the doping efficiency in the $R<10$ nm range. In contrast, DFT calculations \cite{Leao08,Rurali09} predict that $E_b$ decreases much faster than $1/R$, and is about $3-4$ times lower than found in Refs.~\cite{Diarra07,Li08}. In this Letter, we show that present DFT approaches, based on the Kohn-Sham (KS) hamiltonian of the neutral donor, can not predict $E_b$ correctly in bulk and nanostructures, because they miss most of the interactions of the carriers with the polarization charges of the impurity. We propose an alternative strategy based on the KS hamiltonian of the ionized donor which circumvents this deficiency.

For a donor, $E_b$ is the energy needed to ionize the neutral impurity and bring the electron to the conduction band edge far away. It can be defined as the difference $E_b=I^d(N+1)-A^p(N)$ between the ionization energy $I^d(N+1)$ of the neutral impurity (with $N+1$ electrons) and the affinity $A^p(N)$ of the pristine system (with $N$ electrons and no dopant). Since $A(N)=I(N+1)$, the binding energy can also be computed as an isoelectronic difference of ionization energies, $E_b=I^d(N+1)-I^p(N+1)$, or affinities, $E_b=A^d(N)-A^p(N)$. In practice, the ionization energies and affinities can be computed either as total energy differences $I(N)=E(N-1)-E(N)$ and $A(N)=E(N)-E(N+1)$ \cite{Leao08}, or as ``quasiparticle'' energies \cite{Rurali09}, i.e., as the highest occupied (HOMO) and lowest unoccupied (LUMO) molecular orbital energies. However, the quasiparticle problem should in principle be adressed with many-body perturbation theories (MBPTs) such as the $GW$ approximation, since DFT is known to miss the HOMO-LUMO gap \cite{Hedin69,Onida02,Bruneval09,Lannoo85}. We actually show hereafter that the above definitions of $E_b$ are consistent in the $GW$ approximation, but not in DFT. Using the insight gained from many-body theory, we conjecture that DFT should succeed in the calculation of $E_b$ from the KS LUMO of the ionized impurity, but fails in the calculation of $E_b$ from the KS HOMO of the neutral impurity, due to the lack of explicit {\it screened} exchange in the present functionals. We support these conclusions with DFT calculations on Si NWs.

{\it The binding energy in many-body theory} -- In MBPT, the quasiparticle energies $E_n$ and wave functions $\varphi_n$ of the $N$-electron system are the solutions of the quasiparticle equation:
\begin{eqnarray}
&&-\frac{1}{2}\Delta_{\vec{r}}\varphi_n(\vec{r})+v_{\rm ion}(\vec{r})\varphi_n(\vec{r})+v_{\rm h}(\vec{r})\varphi_n(\vec{r}) \nonumber \\
&&+\int d^3\vec{r}'\,\Sigma_{\rm xc}(\vec{r},\vec{r}',E_n)\varphi_n(\vec{r}')=E_n\varphi_n(\vec{r})\,
\label{eqQP}
\end{eqnarray}
where $v_{\rm ion}(\vec{r})$ is the ionic potential, $v_{\rm h}(\vec{r})=\int d^3\vec{r}'\,\rho(\vec{r}')/|\vec{r}-\vec{r}'|$
is the Hartree potential created by the ground-state electronic density $\rho(\vec{r})$, and $\Sigma_{\rm xc}(\vec{r},\vec{r}',E_n)$ is the ``self-energy'' that describes exchange and correlation effects. The ionization energy is $I(N)=-E_N$, while the affinity is $A(N)=-E_{N+1}$.

The $GW$ approximation has become the reference for the calculation of the band structure of semiconductors \cite{Hedin69,Onida02}. For illustrative purposes, we shall use hereafter the simpler static COHSEX form (COulomb Hole and Screened EXchange) of the $GW$ self-energy \cite{Hedin69}. $\Sigma_{\rm xc}$ can then be split in two parts $\Sigma_{\rm COH}+\Sigma_{\rm SEX}$:
\begin{subequations}
\begin{eqnarray}
\Sigma_{\rm COH}(\vec{r},\vec{r}')&=&\frac{1}{2}\delta W_N(\vec{r},\vec{r})\delta(\vec{r}-\vec{r}') \label{eqSigCOH} \\
\Sigma_{\rm SEX}(\vec{r},\vec{r}')&=&-W_N(\vec{r},\vec{r}')\gamma(\vec{r},\vec{r}') \label{eqSigSEX} \\
\gamma(\vec{r},\vec{r}')&=&\sum_{n'\in\{\sigma\}}\varphi_{n'}^*(\vec{r})\varphi_{n'}^{\vphantom{*}}(\vec{r}')\,,
\end{eqnarray}
\end{subequations}
where the sum runs over the occupied states with a given spin $\sigma$. $W_N(\vec{r},\vec{r}')$ is the {\it screened} Coulomb interaction, i.e. the total potential created at point $\vec{r}'$ by a test unit charge at point $\vec{r}$ \cite{noteRPA}. It can also be split in the bare potential $v(\vec{r},\vec{r}')=1/|\vec{r}-\vec{r}'|$ created by this test charge, plus the response $\delta W_N(\vec{r},\vec{r}')=W_N(\vec{r},\vec{r}')-v(\vec{r},\vec{r}')$ of the valence electrons. $\Sigma_{\rm SEX}$ has the same functional form as the Hartree-Fock exchange, but with a screened instead of a bare Coulomb interaction. $\Sigma_{\rm COH}(\vec{r},\vec{r}')$ describes the interaction of a carrier at point $\vec{r}$ with the valence electrons which dynamically respond to its motion.

Before adressing the impurity problem, we shall discuss the form of $W_N(\vec{r},\vec{r}')$ in bulk materials and nanowires. In a solid, a test charge $q_t=+1$ at point $\vec{r}$ attracts valence electrons in a small ``cloud'' around (over $\sim$ a bond length). This cloud contains a total charge $q_c=-(1-1/\kappa)$, where $\kappa$ is the static dielectric constant of the material. The electrons are actually dragged from the surface of the system, where they leave an opposite polarization (or ``image'') charge $q_s=-q_c$. In bulk, these image charges are infinitely far away, so that the long-range potential created by the test charge is simply $W_N(\vec{r},\vec{r}')\sim(q_t+q_c)/|\vec{r}-\vec{r}'|=1/(\kappa|\vec{r}-\vec{r}'|)$. In a nanowire, however, the electrons are dragged within a few $R$'s only from $q_t$, so that the transfer of charges from the surface to the cloud becomes shorter and shorter-ranged with decreasing $R$. The screening is therefore reduced by $q_s$ and the potential ultimately tends to $W_N(\vec{r},\vec{r}')\sim1/|\vec{r}-\vec{r}'|$ when $R\to0$ (i.e., the test charge mostly sees vacuum around for small $R$'s). This simple picture is consistent with classical electrostatics (where the surface polarization charges are given by the discontinuity of the electric field), and fully supported by quantum calculations \cite{Delerue03,Trani06}.

As discussed previously, the many-body binding energy of an impurity can be computed as $E_b=A^d(N)-A^p(N)$, the difference between the affinities of the {\it ionized} impurity and pristine systems. They fulfill the equation $H^{p,d}(N)\varphi_{N+1}^{p,d}=-A^{p,d}(N)\varphi_{N+1}^{p,d}$, where $H^p(N)$ and $H^d(N)$ are the respective quasiparticle hamiltonians:
\begin{equation}
H^{p,d}(N)=-\frac{1}{2}\Delta+v_{\rm ion}^{p,d}+v_{\rm h}^{p,d}+\Sigma_{\rm SEX}^{p,d}+\Sigma_{\rm COH}^{p,d}\,.
\end{equation}
The physics of the impurity is most easily brought out from the difference between $H^p(N)$ and $H^d(N)$. On one hand, the extra proton of the ionized impurity is screened by the valence electrons through the Hartree potential $v_{\rm h}^d$. Neglecting short range chemical corrections in a first approximation \cite{notescreening}, we can therefore write:
\begin{equation}
[v_{\rm ion}^d+v_{\rm h}^d]-[v_{\rm ion}^p+v_{\rm h}^p]\simeq -W_N^d(\vec{r}_i,\vec{r})\,,
\label{eqdeltav}
\end{equation}
where $\vec{r}_i$ is the impurity position. On the other hand, we do not expect significant differences between the screened Coulomb interactions $W_N^d$ and $W_N^p$, nor between the one-particle density matrices $\gamma^d$ and $\gamma^p$, except possibly right around the donor and surface (image charges), on length scales much shorter than the Bohr radius of the impurity. Hence, $\Sigma_{\rm SEX}^d\simeq\Sigma_{\rm SEX}^p$, $\Sigma_{\rm COH}^d\simeq\Sigma_{\rm COH}^p$, and:
\begin{equation}
H^d(N)\simeq H^p(N)-W_N(\vec{r}_i,\vec{r})\,.
\label{eqGWA}
\end{equation}
{\it In a first approximation, the quasiparticle hamiltonian of the ionized impurity is the quasiparticle hamiltonian of the pristine system plus the screened Coulomb potential of a unit charge at the impurity position}. This is the usual ``hydrogenic model'' \cite{Kohn55} used in Refs.~\cite{Diarra07,Diarra08,Li08} to calculate $E_b$ in Si NWs.

The electron is therefore bound to the impurity by the screened Coulomb interaction $W_N(\vec{r}_i,\vec{r})$. In bulk silicon, $W_N(\vec{r}_i,\vec{r})\sim 1/[\kappa|\vec{r}_i-\vec{r}|]$ and $E_b\simeq 50$ meV. In a nanowire, however, the electron also interacts with the image charges of the donor. Since the total surface polarization  charge is $q_s=(1-1/\kappa)\gg1/\kappa$, this leads to a large $\propto 1/R$ enhancement of $E_b$ with decreasing $R$ \cite{Diarra07}.

Let us now compute the binding energy $E_b=I^d(N+1)-I^p(N+1)$ from the ionization energy of the neutral impurity. $I^d(N+1)$ and $I^p(N+1)$ fulfill the equation $H^{p,d}(N+1)\varphi_{N+1}^{p,d}=-I^{p,d}(N+1)\varphi_{N+1}^{p,d}$, where, as before, $H^p(N+1)$ and $H^d(N+1)$ are the quasiparticle hamiltonians of the $(N+1)$-electron pristine and impurity systems. In the latter, the HOMO $\varphi_{N+1}^d$ is the occupied bound state of the impurity. The neutral impurity as a whole now introduces a localized perturbation of the pristine system which is screened by the valence electrons. We can therefore write:
\begin{equation}
[v_{\rm ion}^d+v_{\rm h}^d]-[v_{\rm ion}^p+v_{\rm h}^p]\simeq -W_{N+1}^d(\vec{r}_i,\vec{r})+v_b(\vec{r})\,,
\label{eqdeltav2}
\end{equation}
where:
\begin{equation}
v_b(\vec{r})=\int d^3\vec{r}'\,W^d_{N+1}(\vec{r},\vec{r}')|\varphi_{N+1}^d(\vec{r}')|^2
\label{eqvb}
\end{equation}
accounts for the screening of the bound state potential. Assuming again that $W_{N+1}^p\simeq W_{N+1}^d$, and that the valence band wave functions $\varphi_1,...,\varphi_N$ are little affected by the neutral impurity, we further get:
\begin{eqnarray}
&&\Sigma_{\rm SEX}^d(\vec{r},\vec{r}')-\Sigma_{\rm SEX}^p(\vec{r},\vec{r}') \simeq -W_{N+1}(\vec{r},\vec{r}') \nonumber \\  &&\times [\varphi_{N+1}^{d*}(\vec{r})\varphi_{N+1}^d(\vec{r}')-\varphi_{N+1}^{p*}(\vec{r})\varphi_{N+1}^p(\vec{r}')]\,.
\end{eqnarray}
The second term can be neglected in bulk and nanowires where $\varphi_{N+1}^p$ is an extended state. The first term cancels $v_b(\vec{r})$ when applied to the HOMO $\varphi_{N+1}^d$. The effective hamiltonian for the bound electron therefore reads:
\begin{equation}
H^d(N+1)\simeq H^p(N+1)-W_{N+1}(\vec{r}_i,\vec{r})\,.
\label{eqGWI}
\end{equation}

In principle, $I(N+1)=A(N)$ and we should have recovered the same equation as before [Eq.~(\ref{eqGWA})]. Here $H^p(N)$ is however replaced with $H^p(N+1)$ and $W_N$ with $W_{N+1}$. Since $\varphi_{N+1}^p$ is an extended state, $H^p(N+1)$ and $H^p(N)$ also primarily differ by the substitution $W_N\to W_{N+1}$. The appearance of $W_{N+1}$ introduces a residual ``self-correlation'' error in the $GW$ ionization energies \cite{Romaniello09}, which is however expected to be limited in solids. We can therefore conclude that $GW$ provides a consistent description of the binding energies, whether computed from $A^d(N)$ or $I^d(N+1)$.

This paragraph clearly demonstrates the importance of screened exchange in the calculation of $I^d(N+1)$. Screened exchange indeed cancels the unphysical screened interaction of the bound electron with itself which arises from $v_b(\vec{r})$ in Eq. (\ref{eqdeltav2}). $H^d(N+1)$ is therefore the hamiltonian of a {\it charged} system as expected (the bound electron interacts with $N+1$ ionic charges but $N$ electrons). Such spurious self-interactions are a serious issue in self-consistent descriptions of occupied localized states. In this respect, we would like to point out that the Hartree-Fock (HF) {\it bare} exchange $\Sigma_{\rm x}(\vec{r},\vec{r}')=-v(\vec{r},\vec{r}')\gamma(\vec{r},\vec{r}')$ does not properly correct the {\it screened} self-interactions appearing in solids. Following the same lines as before, the HF hamiltonian of the HOMO of the neutral impurity can indeed be written $H^d_{\rm HF}(N+1)\simeq H^p_{\rm HF}(N+1)-W_{N+1}(\vec{r}_i,\vec{r})+v_b^{\rm sr}(\vec{r})$, where:
\begin{equation}
v_b^{\rm sr}(\vec{r})=\int d^3\vec{r}'\,\left[W_{N+1}(\vec{r},\vec{r}')-v(\vec{r},\vec{r}')\right]|\varphi_{N+1}^d(\vec{r}')|^2 \,. \label{eqvbs2}
\end{equation}
$v_b^{\rm sr}(\vec{r})$ is the spurious potential created by the valence electrons in response to the bound state density $|\varphi_{N+1}^d(\vec{r})|^2$, i.e. the potential created by a diffuse charge $\rho_{\rm eff}(\vec{r})\simeq(1-1/\kappa)\times|\varphi_{N+1}^d(\vec{r})|^2$ plus the opposite surface polarization charge $q_s=-(1-1/\kappa)$. These surface polarization charges balance those embedded in the impurity potential $W_{N+1}(\vec{r}_i,\vec{r})$ (equivalent, as discussed before, to the potential of a net charge $1/\kappa$ at $\vec{r}_i$ and $q_s=(1-1/\kappa)$ at the surface). $H^d_{\rm HF}(N+1)$ is therefore approximately equal to the hamiltonian of the pristine system plus the bare Coulomb potential of a unit charge spread around the impurity (the charge $1/\kappa$ at the impurity position plus the diffuse charge $\rho_{\rm eff}$ around). As a consequence, $\rho_{\rm eff}$ plays the role in the HF approximation of an {\it effective} polarization charge, mislocalized within the scale of the Bohr radius instead of the surface. The relative error on $E_b$ should be limited in thin nanowires (where the Bohr radius is comparable to $R$), and maximum in bulk. The implications for hybrid functionals in DFT will be discussed in the next paragraph.

{\it The binding energy in DFT} -- We now discuss the binding energy of the donor within DFT. For the sake of simplicity, we first focus on the local density (LDA) and generalized gradients (GGA) approximations, then address the case of hybrid functionals. In DFT, the ground-state density $\rho(\vec{r})$ of the $N$-electron system is computed from the eigenstates of the Kohn-Sham hamiltonian \cite{Kohn65}:
\begin{equation}
-\frac{1}{2}\Delta_{\vec{r}}\varphi_n(\vec{r})+[v_{\rm ion}+v_{\rm h}+v_{\rm xc}](\vec{r})\varphi_n(\vec{r})=E_n\varphi_n(\vec{r})\,
\label{eqKS}
\end{equation}
where $v_{\rm xc}(\vec{r})$ is the exchange-correlation potential. In LDA and GGA, $v_{\rm xc}(\vec{r})\equiv v_{\rm xc}\left(\rho(\vec{r})\right)$ is a function of the local density $\rho(\vec{r})$ and of its derivatives. DFT is known to underestimate the band gap energy of semiconductors \cite{Lannoo85}. Still, we show below that DFT should succeed in the calculation of the binding energy from the LUMO of the ionized impurity, but that present functionals fail on the neutral impurity.

Let us first compute $E_b=A^d(N)-A^p(N)$ from the LUMOs of the KS hamiltonians $H^p_{\rm KS}(N)$ and $H^d_{\rm KS}(N)$. The previous arguments are also valid in DFT: The extra proton of the donor is screened by the valence electrons, so that Eq.~(\ref{eqdeltav}) still holds. We do not, moreover, expect much differences between $v_{\rm xc}^{p}(\vec{r})$ and $v_{\rm xc}^{d}(\vec{r})$, except possibly right around the impurity. Therefore, in a first approximation:
\begin{equation}
H^d_{\rm KS}(N)\simeq H^p_{\rm KS}(N)-W_N(\vec{r}_i,\vec{r})\,.
\end{equation}
We hence recover the hydrogenic model as before [Eq. (\ref{eqGWA})]. The KS hamiltonian of the ionized impurity thus embeds the same extra physics (with respect to the hamiltonian of the pristine system) as the $GW$ approximation: {\it Although the LUMO energies are typically underestimated by DFT, the binding energies computed as the difference between the KS LUMOs of the ionized impurity and pristine systems should be reasonably accurate}. This only holds, of course, as long as the binding energy is not too large with respect to the DFT band gap.

Let us now compute $E_b=I^d(N+1)-I^p(N+1)$ from the HOMOs of the KS hamiltonians $H^p_{\rm KS}(N+1)$ and $H^d_{\rm KS}(N+1)$. The KS wave function $\varphi_{N+1}^d$ is the occupied bound state of the impurity. The neutral impurity as a whole is again screened by the valence electrons [Eq.~(\ref{eqdeltav2})]. The exchange-correlation potential $v_{\rm xc}(\vec{r})$ is also affected by the extra bound state density around the impurity. We hence get :
\begin{eqnarray}
H^d_{\rm KS}(N+1)&\simeq&H^p_{\rm KS}(N+1)-W_{N+1}(\vec{r}_i,\vec{r}) \nonumber \\
&+&v_b(\vec{r})+\Delta v_{\rm xc}(\vec{r})\,,
\end{eqnarray}
where $\Delta v_{\rm xc}(\vec{r})=v_{\rm xc}^d(\vec{r})-v_{\rm xc}^p(\vec{r})$. At variance with the $GW$ approximation, $\Delta v_{\rm xc}(\vec{r})$, a local density correction within the Bohr radius, can not be expected to cancel $v_b(\vec{r})$ [Eq.~(\ref{eqvb})], a long-range Coulomb term
. This results in {\it i}) a self-interaction error, and {\it ii}), an almost complete cancellation of the interaction of the electron with the image charges of the impurity. Indeed, $W_{N+1}(\vec{r}_i,\vec{r})$ and $v_b(\vec{r})$ are the potentials created by two opposite charges (the ionized impurity and bound electron), leaving no net charge in the hamiltonian. Both errors give rise to an increase of the impurity level and to a decrease of the binding energy. Although this is especially sensitive in thin nanowires, where the enhancement of $E_b$ is mostly due to the interaction with the image charges, the LDA and GGA would fail up to the bulk [where the impurity potential decreases exponentially instead of $1/(\kappa|\vec{r}-\vec{r}_i|)$]. We stress that the calculation of the ionization energy or affinity of the impurity as a difference of total energies \cite{Leao08}, which involves the neutral impurity as the initial or final state, suffers from the same deficiencies in the LDA or GGA. 

{\it Application to Si nanowires} -- The binding energies of dopant impurities in Si NWs have been previously computed from the KS HOMO of the neutral impurity using GGA and a hybrid functional (HGH) \cite{Rurali09}, i.e. a mixture of Hartree-Fock bare exchange with GGA. As discussed previously, bare echange does not localize the polarization charges properly, the error being however likely limited in thin nanowires (the total charge being correct). The GGA results of Ref.~\cite{Rurali09} are therefore expected to completely miss image charge effects, while the HGH results, which include $12\%$ bare exchange, are expected to account for $\simeq12\%$ of the interactions with the image charges (even though mislocalized). As a consequence, the difference between the GGA and HGH results of Ref.~\cite{Rurali09} should be approximately $12\%$ of the image charge correction given by Eq.~(3) of Ref.~\cite{Diarra07}, that is, $0.12$ eV for $R=1$ nm, $0.17$ eV for $R=0.75$ nm, and $0.25$ eV for $R=0.5$ nm. This is actually in good agreement with the data of Table I of Ref.~\cite{Rurali09}.

To further support the above conclusions, we have computed the binding energy of a P impurity at the center of a hydrogen passivated, $[110]$-oriented Si nanowire with diameter $d=1.73$ nm, either as $E_b^I=I^d(N+1)-A^p(N)$, or as $E_b^A=A^d(N)-A^p(N)$, using KS HOMOs and LUMOs as ionization energies and affinities. The LDA was used in a wavelet basis set as implemented in the BigDFT code \cite{Genovese08}. The neutral impurity was first relaxed in a 660 atoms supercell. Since the treatment of a charged system is still problematic within such a supercell approach, $A^d(N)$ (as well as $I^d(N+1)$ and $A^p(N)$ for consistency) were actually computed in {\it finite} rods with lengths $l$ up to $9.2$ nm ($1584$ atoms). These rods were build from the original 660 atoms supercell by connecting segments of pristine nanowires and hydrogen passivated ends. The binding energies computed from the charged and neutral impurities are respectively $E_b^A=0.93$ eV and $E_b^I=0.06$ eV for $l=9.2$ nm. As expected, $E_b^A$ is much larger than $E_b^I$, and in good agreement with the semi-empirical model of Ref.~\cite{Diarra07} ($E_b=0.92$ eV when $l\to\infty$). This confirms that present functionals are able to predict the binding energies of impurities or defects {\it from the KS hamiltonian of the charged defect}.

To conclude, we have shown, by a formal comparison with the $GW$ approximation, that the donor binding energies computed from the Kohn-Sham hamiltonians of {\it neutral} impurities can be strongly underestimated in semiconductor nanostructures (even with hybrid functionals). This is due to the lack of {\it screened} exchange in the present functionals, and explains the discrepancies between Refs.~\cite{Leao08,Rurali09} and previous works \cite{Diarra07,Li08}. The binding energy of a donor should preferably be computed as the difference between the Kohn-Sham LUMOs of the {\it ionized} impurity and pristine systems. This provides a reasonable substitute for much more expensive $GW$ calculations of defect bound states in solids.

We thank L. Wirtz, X. Blase and H. Mera for fruitful discussions. This work was supported by the french ANR project ``QuantaMonde'' (contract ANR-07-NANO-023-02). The calculations were run at the CCRT and CINES.


\begin{thebibliography}{99}
\bibitem{Bjork09} M.~T.~Bj\"ork {\it et al}., Nature Nanotechnology {\bf 4}, 103 (2009).
\bibitem{Yoon09} J.~Yoon {\it et al}., Appl. Phys. Lett. {\bf 94}, 142102 (2009).
\bibitem{Hedin69} L.~Hedin and S.~Lundqvist, \emph{Solid State Physics} {\bf 23}, ed. by H.~Ehrenreich, F.~Seitz, and D.~Turnbull (Academic Press, New York, London 1969).
\bibitem{Onida02} G.~Onida, L.~Reining, and A.~Rubio, Rev. Mod. Phys. {\bf 74}, 601 (2002).
\bibitem{Kohn55} W.~Kohn and J.~M.~Luttinger, Phys. Rev. {\bf 98}, 915 (1955).
\bibitem{Martins02} A.~S.~Martins {\it et al}., Phys. Rev. B {\bf 65}, 245205 (2002).
\bibitem{Diarra07} M.~Diarra {\it et al}., Phys. Rev. B {\bf 75}, 045301 (2007).
\bibitem{Diarra08} M.~Diarra {\it et al}., J. Appl. Phys. {\bf 103}, 073703 (2008).
\bibitem{Li08} B.~Li {\it et al}., Phys. Rev. B {\bf 77}, 115335 (2008).
\bibitem{Leao08} C.~R.~Leao, A.~Fazzio, and A.~J.~R.~da~Silva, Nano Lett. {\bf 8}, 1866 (2008).
\bibitem{Rurali09} R.~Rurali {\it et al}., Phys. Rev. B {\bf 79}, 115303 (2009).
\bibitem{Bruneval09} F. Bruneval, Phys. Rev. Lett. {\bf 103}, 176403 (2009).
\bibitem{Lannoo85} M.~Lannoo, M.~Schl\"{u}ter, and L.~J.~Sham, Phys. Rev. B {\bf 32}, 3890 (1985).
\bibitem{noteRPA} More precisely, the screened coulomb interaction in the random phase approximation (RPA).
\bibitem{Delerue03} C.~Delerue, M.~Lannoo, and G.~Allan, Phys. Rev. B {\bf 68}, 115411 (2003).
\bibitem{Trani06} F.~Trani {\it et al}., Phys. Rev. B {\bf 73}, 245430 (2006).
\bibitem{notescreening} We also neglect non-linear and exchange-correlation effects beyond the RPA in Eq.~(\ref{eqdeltav}). They would not change the conclusions drawn in this letter.
\bibitem{Romaniello09} P. Romaniello, S. Guyot and L. Reining, J. Chem. Phys. {\bf 131}, 154111 (2009).
\bibitem{Kohn65} W.~Kohn and L.~J.~Sham, Phys. Rev. {\bf 140}, 1133 (1965).
\bibitem{Genovese08} L.~Genovese {\it et al}., J. Chem. Phys. {\bf 129}, 014109 (2008) 
\end{thebibliography}
\end{document}